
\input phyzzx
\baselineskip=18pt
\FRONTPAGE
\hfill {CU-TP-652}\break
\strut\hfill {NI-94-013} \break
\strut\hfill {cond-mat/9409046 }
\vglue 0.8in
\font\twelvebf=cmbx12 scaled\magstep2

\vskip 0.1in
\centerline {\twelvebf Vortices and sound waves  in superfluids }

\vskip .6in
\centerline{\it  Kimyeong Lee$^*$ }
\vskip .05in
\centerline{ Physics Department, Columbia University}
\centerline{ New York, N.Y. 10027, U.S.A.$^\dagger$}

\centerline{and }

\centerline{ Issac Newton Institute for Mathemetical Sciences}
\centerline{University of Cambridge, 20 Clarkson Rd., Cambridge CB3 0EH}
\vskip 0.6in

  We consider the dynamics of vortex strings and sound waves in
superfluids in the phenomenological Landau-Ginzburg  equation.  We
first derive the vortex equation where the velocity of a vortex is
determined by the average fluid velocity and the density gradient near
the vortex.  We then derive the effective action for vortex
strings and sound waves by the dual formulation. The effective action
might be useful in calculating the emission rate of sound waves by
moving vortex strings.
\vfill

\noindent September 1994
\footnote{}{$*$Electronic address: klee@cuphyh.phys.columbia.edu}
\footnote{}{$\dagger$ Current address}

\vfill
\endpage


\def\pr#1#2#3{Phys. Rev. {\bf D#1}, #2 (19#3)}
\def\prl#1#2#3{Phys. Rev. Lett. {\bf #1}, #2 (19#3)}

\def\np#1#2#3{Nucl. Phys. {\bf B#1}, #2 (19#3)}

\def\vR{ {\bf  R}}
\def\vN{{\bf  N}}
\def\vx{{\bf  x}}
\def\vn{{\bf \nabla}}

\REF\rDonnelly{ R.J. Donnelly, {\it Quantized Vortices in Helium II},
Cambridge, Cambridge Univ. Press (1991); R.J. Donnelly, Annu. Rev.
Fluid. Mech. {\bf 25}, 325 (1993); D.R. Tilley and J. Tilley, {\it
Superfluidity and Superconductivity}, Adam Hilger Ltd., Bristol (third
ed.) (1990).}

\REF\r  {M. Hatsuda, S. Yahikozawa, P. Ao and D.J. Thouless, Phys.
Rev. {\bf B 49}, 15870 (1994); M. Sato and S. Yahikozawa, {\it
Topological Formulation of Effective Vortex Strings}, Kyoto Univ.
Preprint, KUNS-1269, HE(TH)94/08, hep-th/9406208.  }

\REF\rBen{U. Ben-Ya'acov, \np{382}{597}{92}; J. Dziarmaga,
\pr{48}{3809}{93}.}

\REF\rGinzburg{V.L. Ginzburg and L.P. Pitaevskii, Sov. Phys. JETP
{\bf 7}, 858 (1958); P. Pitaevskii, Sov. Phys. JETP {\bf 13},
451 (1961);  E.P. Gross, Nuovo Cim., {\bf 20}, 454 (1961);
E.P. Gross, J. Math. Phys. {\bf 4}, 195 (1963).}

\REF\rDHLee{ D.-H. Lee and M.P. Fisher, Int. Jour. of Mod. Phys.
{\bf B5}, 2675 (1991); K. Lee, \pr{48}{2493}{93}; .}

\REF\rFetter{A.L. Fetter, Phys. Rev. {\bf 100}, 1966; M. Rasetti and
T. Regge, Physica {\bf 80 A}, 217 (1975); R.J. Creswick and H.L.
Morrison, Phys. Lett. {\bf A76}, 267 (1980); H.  Kuratsuji,
\prl{68}{1746}{92}.}

\REF\rKalb{M. Kalb and P. Ramond, \pr{9}{2273}{74};
F. Lund and T. Regge, \pr{14}{1524}{76};
R.L. Davis and E.P.S. Shellard, \prl{63}{2021}{89}; R.A. Battaye and
E.P.S. Shellard, \np{423}{260}{94}.}

\REF\rKlee{ Y.B. Kim and K. Lee, \pr{49}{2041}{94}; K. Lee,
\pr{49}{4265}{94}.}

\REF\rArms{ R.J. Arms and F.R.  Hama, Phys. Fluid {\bf 8}, 553 (1965);
G.K. Batchelor, {\it An Introduction to Fluid Dynamics}, Cambridge
Univ. Press, Cambridge (1967).}

\vfill
\endpage

Recently there was some new interest in understanding the dynamics of
vortex strings and sound waves in superfluids or superconductors
[1,2].  While many aspects of the vortex dynamics has been analyzed
extensively and used to many applications, we feel that some aspects
of the basic vortex dynamics, like the vortex equation or the
interaction between vortices and sound waves, are not well understood
as one hoped.  Based on recent progress in rigorous understanding of the
vortex dynamics in relativistic field theoretic models [3], where the
Magnus force on the vortex string is important, we derive hopefully a
new vortex equation.  While this is not a `dynamical' equation of
motion, it leads to some insight into the vortex dynamics. The model
we use to describe superfluids and superconductors is the nonlinear
Schr\"odinger equation studied before [4]. In addition, following the
dual formulation of the theory of a complex scalar field [5], we
derive an effective action for vortices interacting with sound waves.
While many aspects of our goal have been studied before, there still
seems to be a room for improvements from our point of view.

In classical hydrodynamics a vortex moves with the local velocity of
the fluid. Such a view has been applied to vortices in superfluids
[1]. The quantization of the vortex strings has been also investigated
[6].  However, the vortex equation we will derive has an additional
term which depending on the spatial dependence of the fluid density
and has an interesting physical  interpretation.  The vortex string
equation we derive can be regarded as a nonrelativistic limit of the
relativistic string equation.  The relativistic vortex string
equation has been useful in understanding the interaction between
vortex strings very close to each other [3].

Since the work by Kalb and Ramond, there has been many works relating
the antisymmetric tensor field coupled to vortex strings to vortex
strings in superfluids and superconductors.  However, the
antisymmetric tensor field have usually described massless Goldstone
bosons moving at speed of light rather than sound waves of speed less
that unity [7]. The Kalb-Ramond antisymmetric tensor field is
generated by the dual formulation of the massless Goldstone field. The
nonzero fluid density appears as a uniform `magnetic' field for vortex
strings and give rise to the Magnus force on strings. Here we use the
dual formulation to get the effective action for vortex strings and
sound waves, extending the two dimensional result in the first paper
of Ref. [5].  Recently, there has been also some interesting progress
in precise understanding of relativistic vortex dynamics where the
Magnus force is important [8].

We start with the nonrelativistic Lagrangian for a complex scalar
field, which may be coupled to the electromagnetic field,
$$ {\cal L} = {i\over 2}(  \Psi^*\dot{\Psi} - \dot{\Psi}^*\Psi)
- {1\over 2m } |(\vn +ie{\bf A})\Psi|^2 -
{g \over 2} (|\Psi|^2 -\rho_0)^2  + ....
\eqno\eq $$
where the dots denote terms not important for our discussions like
$ eA_0 |\Psi|^2$.  The average charge density is given by a constant
$\rho_0$.  The field equation is then
$$ i\dot{\Psi} = -{1\over 2m} (\vn +ie{\bf A})^2 \Psi + ...
\eqno\eq $$
There is a global Abelian symmetry whose charge density
 and current  are  $\rho = |\Psi|^2$ and
$$ {\bf J}  = -{i\over 2m}\bigg[ \Psi^*(\vn +ie{\bf A})\Psi - (\vn + ie{\bf
A})\Psi^* \Psi \bigg]
\eqno\eq $$
The conservation law would be $\dot{\rho} + \vn \cdot {\bf J} = 0$.

There are quantized vortex strings in the system. With $\Psi=
\sqrt{\rho} e^{i\theta}$, the phase $\theta$ changes by $2\pi$ around
each vortex string. The natural core size of vortex strings would be
given by the coherence length $ \xi =1/\sqrt{ g m \rho_0}$.  To
understand the vortex string dynamics, let us first consider a single
vortex string whose position is given by $\vR(t,\sigma)$ with string
coordinate $\sigma$.  The tangent vector along the string would be
then $\vR'(t,\sigma) = \partial \vR/\partial \sigma $.  We choose the
$\sigma$ sign so that the $\theta$ field increases with the right-hand
rule.  We introduce two unit vectors $\vN_1(t,\sigma), \vN_2(t,\sigma)
$ perpenducular to $\vR'$ At the point on the vortex line, $\vR',
\vN_1,\vN_2$ would form a right-handed orthogonal basis for three
space. There is a new coordinate system $(\sigma, \zeta^1,\zeta^2)$ on
this basis near that point such that
$$ \vx  = \vR(t,\sigma(t,\vx)) +
\sum_{A=1,2} \zeta^A (t,\vec{x})\vN_A(t,\sigma(t,\vx))
\eqno\eq $$
Near the vortex, the relation between $\vx $ and $(\sigma, \zeta^1,
\zeta^2) $ would be invertable.

By evaluating the time derivative of Eq.(4), we get that on the point
on the vortex
$$ \dot{\zeta}_A(t, \vR) = - \dot{\vR}\cdot \vN_A
\eqno\eq $$
Similarly, the space derivative  $\partial/\partial x^j$ on Eq.(4) leads
to
$$\eqalign{ &\ \vn \sigma = { \vR' \over |\vR'|^2}  \cr
   &\ \vn\zeta^A = \vN_A \cr}
   \eqno\eq $$
on the  point. Two spatial derivative $\vn^2 $ of
Eq.(4) leads to
$$ \vn^2 \zeta^A = -{ \vR'' \cdot \vN_A  \over |\vR'|^2}
\eqno\eq $$

To obtain a vortex equation from Eq.(1), we first  notice that
near a  vortex string,  the scalar field would be $\Psi
\sim \psi_s \equiv (\zeta^1 + i \zeta^2)$.  In general the scalar
field can be written as $\Psi = \psi_s\Phi$, where $\Phi$ would not in
general vanish on the vortex string.  Let us evaluate Eq.(1) on a
point on the string, to get
$$ i \Phi\dot{\vR}\cdot \vn\psi_s
= -{1\over 2m} \biggl[ \Phi\vn^2 \psi_s + 2\vn \Phi \cdot
\vn \psi_s + 2i e \Phi {\bf A} \cdot\vn \psi_s ) \biggr]
\eqno\eq $$
The rest of terms vanishes on the vortex string. After dividing Eq.(8)
by $\Phi$ and  using Eqs.(5), (6) and (7),
we get
$$
-i (\vN_1 + i\vN_2 )  \cdot \dot{\vR}  = -
{1\over 2m} (\vN_1 + i \vN_2) \cdot \left\{ - { \vR'' \over |\vR'|^2}
+ 2\vn\ln |\Phi| + 2i \bigl( \vn {\rm Arg}\, \Phi
+ e{\bf A}  \bigr) \right\}
\eqno\eq $$
Identifying the real and imaginary parts separately, we get the
equation of motion for vortex strings,
$$ \dot{\vR } = \alpha \vR'
- {1\over m} {\vR' \over |\vR'|^2}
\times \biggl( \vR'\times (\vn  {\rm Arg}\,
\Phi + e {\bf A})\biggr)
 - {1\over 2 m} {\vR' \over |\vR'|} \times \biggl( -{\vR'' \over
|\vR'|^2}  + 2\vn \ln| \Phi | \biggr)
\eqno\eq $$
where $\alpha$ is  undetermined. One convenient parameterization is
such that $\dot{\sigma} = 0$ and so $\dot{\vR}\cdot \vR'=0$, in which
case $\alpha$ vanishes. This is the equation which gives the velocity of
a vortex string once the field configuration $\Phi$ is known at a
given moment. However, it is not a dynamical equation of motion in
ususal sense  because  we do not  in general  know the field $\Phi$ in
terms of a given configuration of vortices. Sometimes,  we can
make a good guess of $\Phi$ and will get some insight into the
vortex dynamics.  Furthermore, the above equation might be a good
start for understanding two vortices crossing each other.  This
equation would be a sort of nonrelativistic limit of the vortex string
equation in the relativistic field theory [3], as the term proportional
to $\ddot{\vR}\times \vR'/m$ is negligible in that limit.

Let us study the implications of Eq.(10) on the vortex dynamics.  The
field $\Phi$ would not change much over the scale $\xi$ near a vortex
and would be determined at a point near a vortex by the shape of the
vortex and other vortices and by other factors. The local fluid
velocity near the vortex would be a sum of the average induced
velocity and the current which goes around the vortex string.
Naturally, the vortex motion would be determined by the average
induced velocity,
$$ {\bf u} = {1\over m} (\vn {\rm Arg} \, \Phi + e{\bf A} )
\eqno\eq $$
In Eq.(10), the second term in the right hand side would lead to the
velocity of the vortex string  at given point is $\dot{\vR} = {\bf u}$, if we
neglect the last term. This is the standard equation for vortices in
incompressible fluids. The above expression for the average velocity
near the vortex is clearly better than the naive expression, $(\vn \theta
+ e{\bf A})/m$ which is very singular at the vortex position.

The term proportional to $\vR'\times \vR''$ is interesting.
The self-induction contribition to the fluid velocity (11) near the
vortex, as we will see, has such a term with much larger coefficient.
Thus we can regard the  term proportional to $\vR'\times \vR''$
as  a small correction to the induced velocity near the vortex.

The last term in Eq.(10) is relevant when the charge density has a
spatial dependence, which is possible because the fluid is
compressible.  To understand its physics concretely, let us consider a
vortex effectively moving on a two dimensional (x,y) plane where the
charge density is forced to decrease along a direction, say along the
$+x$ direction. We are imagining here, for example, a three
dimensional fluid on a container whose bottom is not leveled and so
the effective two dimensional density decreases along the $+x$ axis,
and we assume a vortex is straight along the vertical direction.
Eq.(10) tells us that it would move along the $+y$ direction. The
physics behind this motion is clear. The fluid rotation around the
vortex would introduce more fluid to the $-y$ direction from the $-x$
direction and takes out more fluid from $+y$ direction to the $-x$
direction. Thus, the fluid pressure difference would push the vortex
to the $+y$ direction.  However, we can expect the net fluid velocity
is zero far way from vortex as the above phenomena is simply a
continuous exchange of position among the vortex and the neighboring
fluid. This effect can in principle be tested in a experiment
of a rotating superfluid in a bucket whose bottom is tilted.
 This phenomena looks somewhat similar to an ascending oil drop
in water by buoyancy.

Since our fluid is compressible, there are sound waves due to the
fluctuations, $\Psi = \sqrt{\rho_0 + \delta \rho} e^{i\theta}$.
Linearized equation for the fluctuation implies that $\delta \rho =
- \delta \dot{\delta \theta}/g $ and the sound speed is
$$ v_s = \sqrt{ g \rho_0 \over m}
\eqno\eq $$
{}From Eq.(10), we see the coupling of sound waves to vortex strings is
primarily due the fluid current fluctuations $\delta \theta$ in the
long-wave length limit. The coupling via the density fluctuation term
would become weaker by inverse of the wave length.

To understand the interaction between sound wave and vortices, it is best
to use the dual formulation of the original theory. Since the dual
formulation has been derived many times in past [5], we will summerize
briefly. We start with variables $\rho, \theta$ with the measure
$[d\rho d\theta]$.
We introduce  an auxilliary vector field, ${\bf I}$ such that
$$ \int [d{\bf I}] \exp \biggl\{i \int d^4x \left( { m \over
2\rho } {\bf I}^2 - {\bf I}\cdot \vn \theta \right)\biggr\}
= \rho^{3/2} \exp \biggl\{ i \int d^4x \,\,  {\rho \over 2m} (\vn
\theta)^2  \biggr\}
\eqno\eq $$
where irrelevant numerical factors are dropped. The phase variable is
given as a sum of the singular part $\bar{\theta}$ and the nonsingular
single-valued part $\eta$.  Because $[d\theta]= [d\bar{\theta} d\eta]$,
we can now integrate over the single valued $\eta$, getting the charge
conservation constraint $\dot{\rho} + \vn \cdot {\bf I} = 0 $.
This can be solved by introducing the antisymmetric tensor field
$B_{\mu\nu}$ such that $(\rho, {\bf I})^\mu =
\epsilon^{\mu\nu\rho\sigma} \partial_\nu B_{\rho\sigma}/2$. The dual
Lagrangian would be then written in terms of $B_{\mu\nu}$ and
$\bar{\theta}$.

The coupling between vortices and $B_{\mu\nu}$
would be then $\epsilon^{\mu\nu\rho\sigma}\partial_\mu\bar{\theta}
\partial_\nu B_{\rho\sigma}/2$. This can be written better by
introducing the topologically conserved antisymmetric tensor current
for vortices,
$$ \eqalign{ K^{\mu\nu} &\ = \epsilon^{\mu\nu\rho\sigma}
\partial_\rho\partial_\sigma
\bar{\theta} \cr
&\ = 2\pi \sum_a  \int  d\sigma (\dot{R}_a^\mu R_a'^\nu  - R_a'^\mu
\dot{R}_a^\nu)  \delta^3(\vx - \vR_a(t,\sigma) )
\cr}
\eqno\eq $$
where $R_a^\mu = (t, \vR_a(t,\sigma))$ is the position for the a-th
vortex.  This current is conserved identically, $\partial_\mu
K^{\mu\nu} = 0 $ and leads to a conservation charge $\int_S d \Sigma^i
\cdot K^{0i}$ on any surface $S$ without boundary,  which counts the
net number of vortices piercing the surface.  Since the charge density
is near constant outside the vortex string core region of the scale
$\xi$, the current conservation becomes $\partial_i^2 \theta=0$ in
that outside region.  Then, we can use Eq.(14) to get the Biot-Savart
law
$$ \partial_i \theta = {1\over 2} \sum_a \oint d \vR_a \times{  (\vx - \vR_a)
\over |\vx -\vR_a|^3}
\eqno\eq $$
We emphasize that the solution is valid only outside the vortex core
region. From this one can get ${\rm Arg} \, \Phi$ near a vortex and
can calculate the induced speed ${\bf u}$. It is well known [9] that
the above expression for a single vortex can be approximated by
$$ \partial_i \theta (t,\zeta^A , \sigma) = { \vR' \times \zeta^A \vN_A
\over  |\vR'| \zeta^B \zeta^B } + {1 \over 2}  { \vR' \times
\vR'' \over |\vR'|^3   } \ln (L/\xi)
\eqno\eq $$
near a vortex position. Here the large length cut off is given as $L$
and the current is calculated at the point $\zeta^A \zeta^A \approx
\xi^2$.  The first term in the right-hand side would be the current
around the vortex string. The second term is $\partial_i{\rm Arg } \,\Phi$ and
is fixed by the local self-induced current.  The second term is in
general much bigger than the $\vR'\times \vR''$ term in Eq.(10).
Thus Eq.(16) gives us the $\Phi$ field outside the vortex core region.
Then Eq.(10) allows us to calculate the vortex velocity. This is how
the cutoff $\xi$ comes in the calcuration naturally.  If we
have considered magnetic flux vortices in an extremal type II
superconductor, the natural value for the cutoff $L$ would be the
penetration length $\lambda$ which is much longer than the coherence
length $\xi$.

The field strength of $B_{\mu\nu}$ is $H_{\nu\rho\sigma} =
\partial_\nu B_{\rho\sigma}+ \partial_\rho B_{\sigma\nu} +
\partial_\sigma B_{\nu\rho} $.  The dual Lagrangian becomes
$$ {\cal L}_D = {m H_{0ij}^2 \over 4H_{123} }- { (\vn
H_{123})^2 \over 8mH_{123}} - { g \over 2}(H_{123} - \rho_0)^2 +
{1\over 2}B_{\mu\nu}K^{\mu\nu}
\eqno\eq $$
The measure for the path integral changes from $[d\rho d\theta]$ to $[
H_{123}^{-3/2} dB_{\mu\nu} d\vR_a ]$. There will be also the gauge
fixing term for the antisymmetric tensor field.  (If there is a gauge
coupling, we can integrate over the gauge field $A_\mu$, getting
additional terms in the dual Lagrangian.)  The above expression is
exact. Due to the second term in Eq.(17), the energy of a vortex does
not diverge at the vortex position only if the charge density vanishes
at the position at the vortex. Naively, there seems that a  vortex
equation would arise from the above dual Lagrangian under the
variation of the vortex position $\delta \vR$.  However, the
coefficient $H_{\mu\nu\rho}$ of the variation vanishes at the vortex
position and so there is no additional equation. The reason behind is
of course that the dynamics of vortices is completely determined by
the field dynamics around vortices.  Classically, the antisymmetric
tensor field is related to the original current by
$$ J^\mu = {1\over 6} \epsilon^{\mu\nu\rho\sigma} H_{\nu\rho\sigma}
\eqno\eq $$

In the absence of vortices, the energy of the system would be minimum
when the expectation of $H_{123} $ is the background  charge density
$\rho_0$. The small fluctuations  around this background are  sound
waves and described by a simple Lagrangian
$$ {\cal L}_s = {m \over 4\rho_0} H_{0ij}^2 - { g \over 2}
H_{123}^2 \eqno\eq $$
in the long wave length limit. Here we have shifted the antisymmetric
tensor field

When we consider the small perturbations on the field configuration of
moving vortices, we can regard the charge density $H_{123}=\rho$ to be
almost constant outside the vortex string core region of the  transverse
radius $\xi$.  With the cutoff near vortices of length scale $\xi$,
the dual Lagrangian can be written as
$$ {\cal L}_D = { m\over 4\rho_0 }H_{0ij}^2 - { g \over
2}(H_{123}-\rho_0)^2 - {1\over 2}B_{\mu\nu}K^{\mu\nu}
\eqno\eq $$
Here we neglect the terms higher order in spatial derivatives or  density
fluctuations.  This is the effective Lagrangian for vortex strings and
the sound wave. We can go further by choosing the gauge for
$B_{\mu\nu}$ so that $\partial^i B_{ij}=0$ and $ \partial_i B_{0i} = 0
$. First, there will be decoupling between the $B_{0i}$ and $B_{ij}$
in the Lagrangian. We can now integrate over the auxiliary variables
$B_{0i}$. The variation of $B_{0i}$ leads to the Gauss law
$$  {m \over \rho_0} \partial_j^2 B_{0i} = 2\pi \sum_a \int d R^i_a
\delta(\vx - \vR_a)
\eqno\eq   $$
We substitute $B_{0i}$ satisfying Eq.(21) to Eq.(20). We shift the spatial
components of the antisymmetric tensor field as
$$B_{ij} = {1\over 3}\rho_0\epsilon_{ijk}x^k  + \sqrt{\rho_0\over m}
\epsilon_{ijk}C^k
\eqno\eq $$
where ${\bf C}$ satisfies the gauge $\vn \times {\bf C}=0$ and can be
regared as the fluctuation around the uniform charge density.

After that we end up with the effective Lagrangian for vortices and
sound waves with a cut off scale $\xi$
$$  \eqalign{ L_{\rm eff} = &\  {2\pi \rho_0 \over 3} \sum_a\oint
d \vR_a \cdot \vR_a \times  \dot{\vR}_a  - \sum_{a,b} {\pi\rho_0 \over m}
\oint\oint {d\vR_a \cdot d\vR_b \over |\vR_a - \vR_b|} \cr
&\ +\int d^3 \vx  \biggl\{ {1\over 2}   \dot{\bf  C}^2
- {v_s^2 \over 2}  (\vn \cdot {\bf C})^2 \biggr\}
+ 2\pi \sqrt{\rho_0 \over m}   \sum_a \oint d\vR_a \cdot {\bf
C}(t,\vR_a)\times \dot{\vR}_a
\cr}
\eqno\eq $$
where the gauge fixing condition $\vn \times {\bf C} =0$ is assumed.
The first two terms has been studied before [6].  The above action is
the effective Lagrangian for vortices interacting with  sound waves.
It has been known that the vortex equation has many interesting time
dependent solutions like, Kelvin waves, Hasimoto solitons, and smoke
rings. Perhaps, the above action will allow us to calculate the
emission rate of sound waves by these solutions. This effective action
might be also useful in understanding the superfluid turbulence.  We
can quantized the reduced effective action by the path integral
formalism or the canonical formalism.  That could allow us to
calculate the energy levels of the excited vortex string loops and the
transition amplitudes between the energy levels by the phonon emission.

Since superfluid is compressible, there could be supersonic shock
waves which  interact with vortex strings. In addition, vortex
strings could move faster than the sound speed, still being
nonrelativistic.  It would be interesting to understand this
supersonic phenomena. The dual action (17) might be useful there.

Finally, it would be interesting to get an effective action for
magnetic flux vortex strings in extreme type II superconductors,
interacting with vector gauge bosons of small mass. This effective
action probably approaches the effective action (23) in the limit where
the electromagnetic coupling constant $e$ goes to zero.

\vskip 0.2in

\centerline{\bf ACKNOWLEDGMENTS }

This work was supported in part by the Department of Energy, the NSF
Presidential Young Investigator program and the Alfred P. Sloan
Foundation. The part of this work was done while the author was
attending the `Topological Defects' workshop in Isaac Newton Institute
for Mathematical Sciences and the author thanks the organizers and
the participants of the workshop for providing an exciting environment.

\endpage

\refout
\end